\documentclass[conference]{IEEEtran}
\IEEEoverridecommandlockouts
\usepackage{cite}
\usepackage{amsmath,amssymb,amsfonts}
\usepackage{algorithmic}
\usepackage{lipsum}
\usepackage{subfig}
\usepackage{graphicx}
\usepackage{textcomp}
\usepackage{xcolor}
\usepackage{url}

\def\BibTeX{{\rm B\kern-.05em{\sc i\kern-.025em b}\kern-.08em
    T\kern-.1667em\lower.7ex\hbox{E}\kern-.125emX}}
\begin{document}

\title{eBPF-based Working Set Size Estimation in Memory Management \\
\thanks{*equal contribution}
}

\author{
\IEEEauthorblockN{Zhilu Lian*}
\IEEEauthorblockA{\textit{School of Software Engineering} \\
\textit{Sun Yat-sen University} \\
Zhuhai, China \\
lianzhlu@mail2.sysu.edu.cn}
\and
\IEEEauthorblockN{Yangzi Li*}
\IEEEauthorblockA{\textit{School of Software Engineering} \\
\textit{Sun Yat-sen University} \\
Zhuhai, China \\
liyz25@mail2.sysu.edu.cn}
\and
\IEEEauthorblockN{Zhixiang Chen}
\IEEEauthorblockA{\textit{School of Software Engineering} \\
\textit{Sun Yat-sen University} \\
Zhuhai, China \\
chenzhx69@mail2.sysu.edu.cn}
\and
\IEEEauthorblockN{Shiwen Shan}
\IEEEauthorblockA{\textit{School of Software Engineering} \\
\textit{Sun Yat-sen University} \\
Zhuhai, China \\
shanshw@mail2.sysu.edu.cn}
\and
\IEEEauthorblockN{Baoxin Han}
\IEEEauthorblockA{\textit{School of Software Engineering} \\
\textit{Sun Yat-sen University} \\
Zhuhai, China \\
hanbx3@mail2.sysu.edu.cn}
\and
\IEEEauthorblockN{Yuxin Su}
\IEEEauthorblockA{\textit{School of Software Engineering} \\
\textit{Sun Yat-sen University} \\
Zhuhai, China \\
suyx35@mail.sysu.edu.cn}
}

\maketitle

\begin{abstract}
Working set size estimation~(WSS) is of great significance to improve the efficiency of program executing and memory arrangement in modern operating systems. 
Previous work proposed several methods to estimate WSS, including self-balloning, Zballoning and so on. However, these methods which are based on virtual machine usually cause a large overhead. Thus, using those methods to estimate WSS is impractical. 
In this paper, we propose a novel framework to efficiently estimate WSS with \textbf{eBPF (extended Berkeley Packet Filter)}, a cutting-edge technology which monitors and filters data by being attached to the kernel. 
With an eBPF program pinned into the kernel, we get the times of page fault and other information of memory allocation. Moreover, we collect WSS via vanilla tool to train a predictive model to complete estimation work with LightGBM, a useful tool which performs well on generating decision trees over continuous value. 
The experimental results illustrate that our framework can estimate WSS precisely with 98.5\% reduction in overhead compared to traditional methods.

\end{abstract}

\begin{IEEEkeywords}
eBPF, Working Set Size Estimation, LightGBM
\end{IEEEkeywords}

\section{\textbf{Introduction}}

With the rapid development of cloud computing services,  data size are soaring and our memory requirements are also skyrocketing. 
But the problem is that the size of memory is limited in every physical machines.  
Therefore, maximizing the utilization of memory is the urgent task to solve the problem of memory shortage.

According to NIST\cite{NIST},  
cloud computing resources includes Iaas (Infrastructure as a Service),  SaaS  (Software as a Service) and PaaS (Platform as a Service). The dispersion of the resources makes cloud computing has wider coverage across the world.  This increases its use especially high-performance needs,  leading to resources management problems especially memory management problems.  Furthermore,  resource over commitment is used for a better utilization of resources of virtualized cloud infrastructure.  Allocating more virtual resources to a machine  (or a group of machines)   than it  (or them) actually can use is called resource over commitment.  Since most applications will not fully/completely use the resources they apply for at all the time,  this approach can save memory for more other users to use so that gains more profit\cite{overCommitment}.  
Difficulties in balancing satisfying of SLAs (Service Level Agreements) and the use of over commitment \cite{SLAs}  
exacerbate these problems.


Therefore, efficient memory management techniques are of great significance for cloud computing.  Among them,  the most common technique is \textbf{Working Set Size (WSS) estimation} techniques.  
WSS is the amount of memory a process needs to keep working during a period of time~\cite{DBLP:journals/pomacs/NituKYTHA18},  i.e. the collections of pages that are actually going to be accessed.  
Therefore, collecting WSS in VMs with non-intrusive and low-overhead method is of great essence.
WSS estimation techniques usually consist of data collecting and data processing.  
The main challenges of WSS estimation come from data collecting,  for the accuracy of estimation mainly depends on the quality of the data it provides.
A great amount of papers have devoted to estimating WSS more accurately and efficiently for it plays a vital role in improving memory management.  
However, previous techniques that used to estimate the WSS have difficulty in balancing efficiency and availability,  which means efficient techniques are more complex to implement, and those techniques that are easier to implement are inferior in performance.  
Moreover, previous methods with high performance and accuracy are not practical in industry, for they work intrusively and thus are unacceptable for most customers~\cite{cassagnes2020rise}.

Among them, constructing MRC~(Miss Ratio Curve) by reuse distance method is the most widely used method.
For the reuse distance method, it takes efforts to lower the space overhead and time overhead due to memory accesses and to balance the overhead and the accuracy of MRC.  
Using internal counters as metric to estimate WSS~\cite{b13},  the method is less persuasive for it doesn't offer what kind of internal counters are being used to get corresponding information.

In opposite to previous methods, we propose a novel method to collect various data in user-space processes, so that we could evaluate WSS more accurately.  
To achieve this objective, we employ \textbf{eBPF (extended Berkeley Packet Filter)} to monitor the execution status of working programs.  
eBPF is the extension of BPF.  
\textit{BPF is a kernel architecture for packer filtering.} 
eBPF transform BPF into a universal inkernel virtual machine. 
The eBPF programs can be written in many languages such as C and Python.
Compared to other data collection tools, eBPF tools show transcendent performance and non-intrusion~\cite{cassagnes2020rise}, enabling dynamic monitoring and tracing of numerous memory activities such as allocating pages, swapping pages and so on with an extremely low overhead.


Specifically, our work can divides into three parts:
\begin{enumerate}
    \item \textbf{Data collecting.}
    As mentioned above,  eBPF is a helpful tool we use to monitor memory allocation. In this paper,  we use Python to write our eBPF monitoring program.  After successful compilation,  the verifier will check whether the program is legal. Then the prepared eBPF programs will be loaded into the kernel.  Attached to an interface called hook, the eBPF programs can operate normally.  Data collected by eBPF are WSS and page faults times during a period of time.  These two objects are  most commonly used in WSS estimation.
    
    \item \textbf{Preprocessing the collected data.}  The deviation between the eigenvalues of the eigenvectors caused by different units of measurement is large,  which is likely to make bad influence on estimation results.  To alleviate this issue,  we use normalization method to process the collected data. 

    \item \textbf{Modeling and estimating.}  We employ processed data to set up and train $LightGBM$ model,  an enduring model in machine learning, to predict WSS more accurately. To reduce over-fitting,  we append penalty terms to our estimation model. It is of great importance for improving the generalization ability of our model.  
\end{enumerate}

The results of our research indicate that the method we use has accurate estimation and excellent performance with non-intrusion, with 0.0744 RMSE in average and a 65x reduction in time overhead.
In summary,  contributions of this paper are as follows: 
\begin{itemize}
    \item To the best of our knowledge, we are the first to estimate the working set size of program execution in non-intrusive way. We employ machine learning method to establish the connection between the count of page fault and the real working set size.
    \item We implement a non-intrusive page fault counting method based on the lightweight observation to Linux kernel with the help of eBPF.
    \item Experimental results illustrate that our proposed framework achieve great accuracy with 0.0744 RMSE in average and high efficiency with 65 times reduction in time overhead to estimate the working set size.
\end{itemize}


\section{\textbf{Background}}
\label{sec:Background}
\subsection{Working Set Size Estimation}

There are a lot of papers which discuss how to calculate the working set size(WSS). Previous work \cite{DBLP:journals/pomacs/NituKYTHA18}\cite{DBLP:journals/jcst/ShaHLWW20} summarize several ways to achieve the goal:
\begin{itemize}
    \item \textbf{self-balloning.} It is a virtual-machine-based way that calculates WSS according to the value of internal counters. 
    \item \textbf{Zballoning.} It estimates WSS by manually adjusting the value of internal counters through different events, and then estimate WSS according to the adjusted value
    \item \textbf{Geiger.}  Geiger. It allocates a small size of memory first. If the allocated memory is insufficient, the ghost buffer takes effect. Therefore, the WSS can be expressed as the equation:
    $$WSS = M_{allocated} + M_{ghost\_buffer}$$
    \item \textbf{LRU-Based MRC Construction.} It draws a picture of MRC according to the LRU stack which can calculate a the reuse distance of a referenced page and estimate WSS through the picture.
\end{itemize}

In this paper, we use the vanilla WSS estimation tool\footnote{https://github.com/brendangregg/wss} to get WSS. 
The tool estimates WSS by resetting a reference flag on memory and checking the number of pages that the flag returns to. We regard the returned WSS as correct value and evaluate our trained model depending on it.

\subsection{Page Fault in Memory Management}

A page fault exception occurs when the working set size~(WSS) dose not meet the requirements of user program.  It is easily understand that when WSS is greater or equal to the memory size requested by a program~\cite{DBLP:journals/access/HarbyFA19}.  Therefore, WSS can be estimated by increasing the memory requested by the program gradually to cause page fault.

In this paper, we collect the numbers of page fault and the time when page fault occurs by executing a program which apply for larger size of memory. We change the size of requested memory to get different data. The specific details will be explained in the section~\ref{sec:Method}.

\subsection{eBPF Technology}

eBPF is the extension of BPF. 
BPF is a kernel architecture for packer filtering~\cite{DBLP:conf/usenix/McCanneJ93}. 
eBPF transform BPF into a universal inkernel virtual machine.
The eBPF programs can be written in many languages such as C and Python. In this paper, we use Python to write our eBPF monitoring program. Figure \ref{fig:2} shows the framework of eBPF. 
The eBPF structure encapsulates 11 registers  with 64 bits, a stack of 521 bytes, a map and some helper functions~\cite{DBLP:journals/csur/VieiraCPSJV20}. 

A few structures support the execution of the eBPF program:

\textbf{(1) Compiler.} As we mentioned before, an eBPF is usually written in C language. It is compiled by the Clang compiler whose back-end is LLVM. Clang compiler compiles the eBPF program according to the rules of C language. However, some functions of C set is not available for eBPF program such as \textit{printf()}.

\textbf{(2) Verifier.} The verifier in the kernel will check the compiled eBPF program before it is loaded.In order to guarantee the security and safety, the verifier will traversal the whole program to see whether it is friendly to the kernel.

\begin{figure}[ht]
\centerline{\includegraphics[scale=0.3]{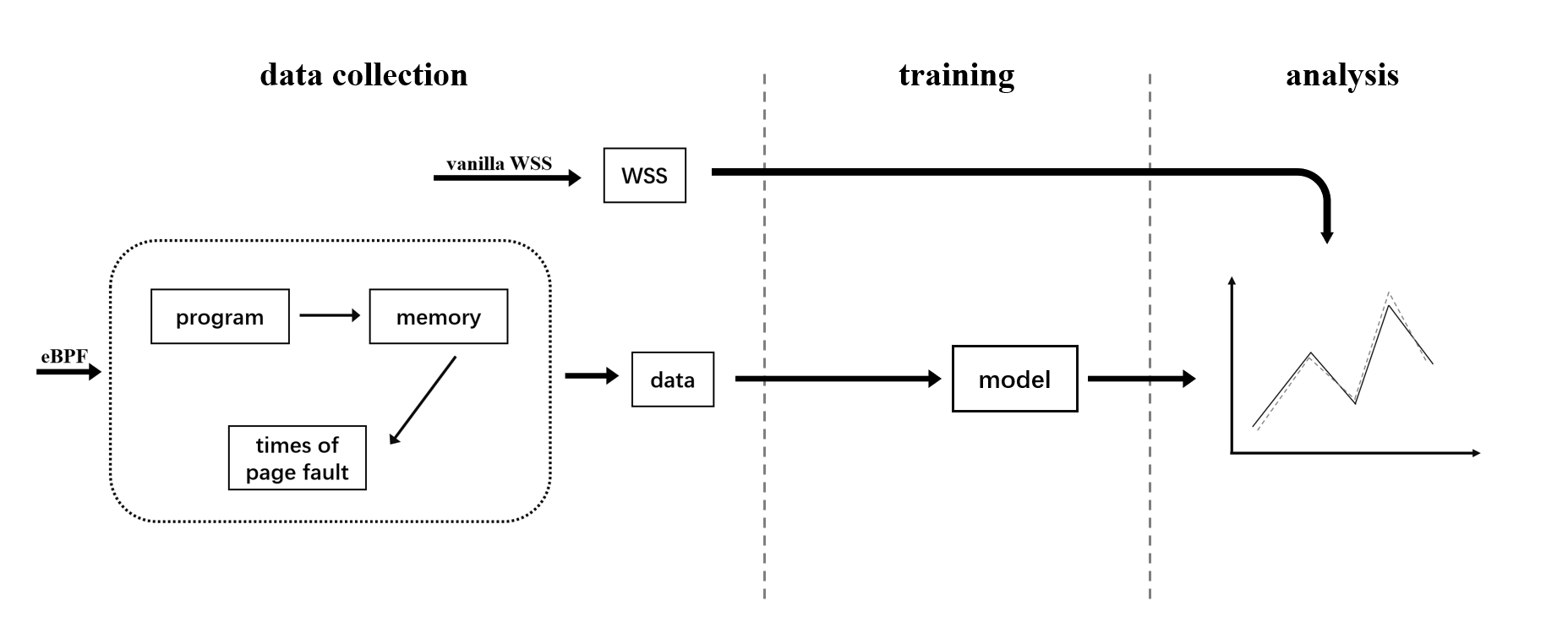}}
\caption{Overall Framework}
\label{fig:1}
\end{figure}

\begin{figure}[ht]
\centerline{\includegraphics[scale=0.3]{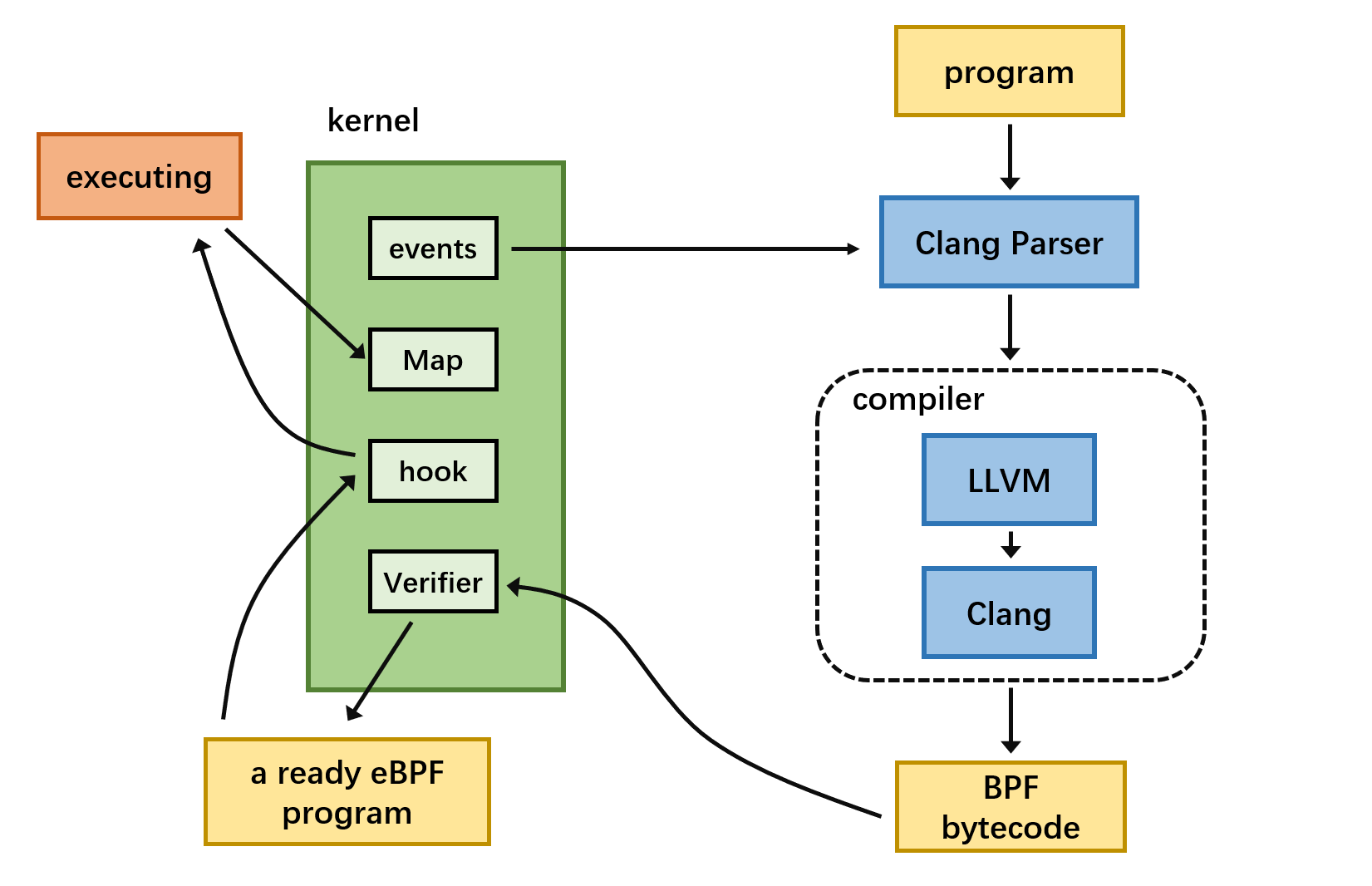}}
\caption{Overall Framework of eBPF}
\label{fig:2}
\end{figure}

\textbf{(3) Maps.} Maps store the key value of eBPF programs. The key value maps the program to system call. An eBPF program should declare a global variable of type \textit{struct bpf\_map\_def} to access a map. Each map consists of a few file descriptors, which will be passed by file loader during loading. Those descriptors will be transformed into pointers by verifier.

\textbf{(4) Helper functions.} The type of an eBPF program not only determines the parameter but also determines what helper functions the program will use and which hook the program attached to. 

\textbf{(5) Tail calls.} The eBPF programs can call other programs by tail calls, which cause low overhead compared to function call and simplify the complicated program. The usage of tail calls involves a specific map called \textit{BPF\_MAP\_TYPE\_PROG\_ARRAY}, which stores references of eBPF programs. Tail calls execute via a helper function called \textit{bpf\_tail\_call}.

Compiled successfully, the verifier will check whether the program is legal. Then the prepared eBPF programs will be loaded into the kernel. With attached to an interface called hook, the eBPF programs can operate normally.

\section{Methodology}
\label{sec:Method}
In this paper, we estimate the size of working set by monitoring the times of page fault. We prepare a few programs that are easy to cause page fault by applying for memory continually and get the data of missing page in memory by eBPF technology. Then we train the WSS estimation model by LightGBM with the data , a tool which can train data into Gradient Boosting Decision Tree~(GBDT). Figure~\ref{fig:1} shows the overall framework.

\subsection{eBPF Inspection}

eBPF programs executed when the corresponding code path is traversed, which is useful network program. In this paper, we apply eBPF technology to monitoring memory allocation. 
We use eBPF technology to collect data. To execute an eBPF program and collect target data, a few works need to be done.

\subsubsection{Source Code}

\begin{figure}[t]
\centerline{\includegraphics[scale=0.4]{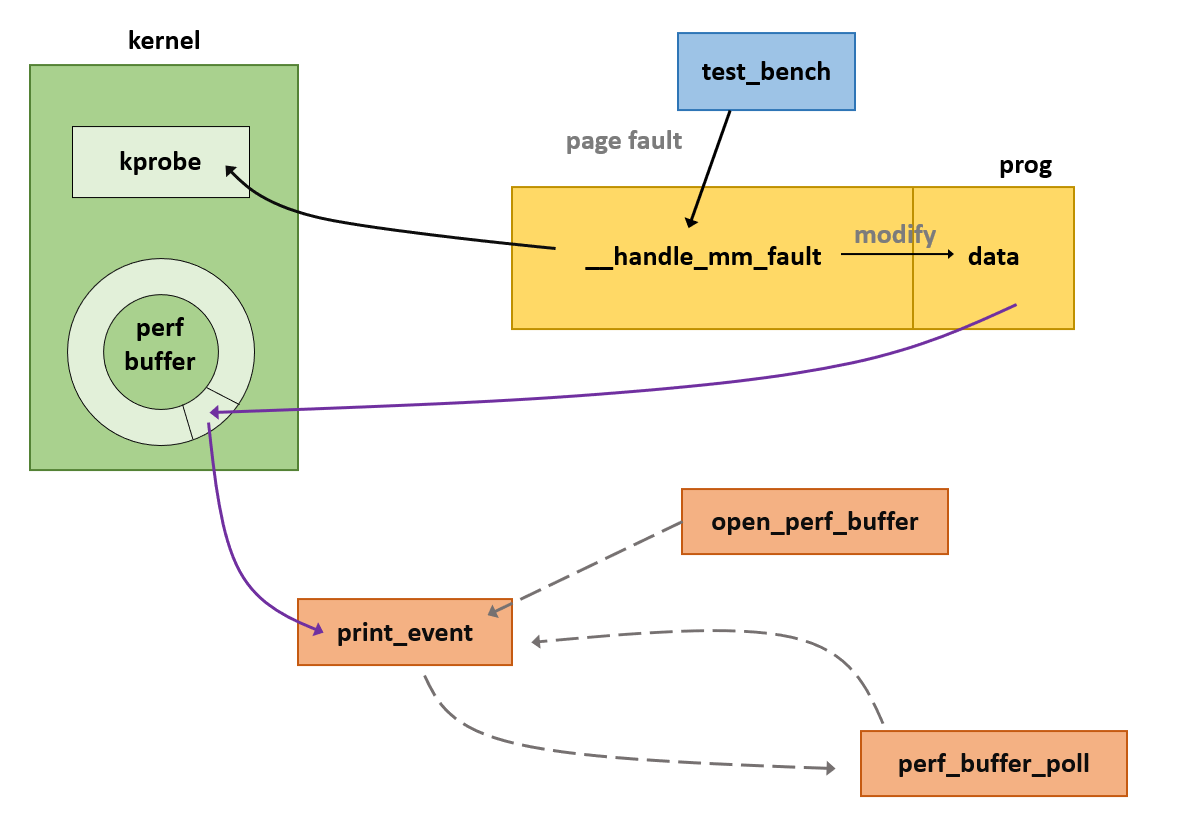}}
\caption{The implementation of our eBPF program}
\label{fig:5}
\end{figure}

In the eBPF program we develop, We use several functions and variables defined in eBPF package. Figure~\ref{fig:5} shows the functions and connections of each part of the eBPF program. 

\begin{itemize}
    \item \textbf{kprobe.} kprobe is the specific event of bpf tools, which can associate the following function with a kernel function by creating dynamic mapping and filtering the data we need.
    \item \textbf{BPF\_HASH(name, key\_type, leaf\_type, size)}. The function defined in the eBPF kernel can create a hash table \textit{name} to store the key value pairs. In this paper, the key value pairs relate the pid to the corresponding times of page fault. In our program, the name is set to \textit{events} and the key\_type is set to u32.
    \item \textbf{BPF\_PERF\_OUTPUT(name)}. The function creates an output table name to deliver event data through a per ring buffer. In our program, the name is set to \textit{events}.
    \item \textbf{handle\_mm\_fault(struct pt\_regs *ctx)}. The function following kprobe records the information of page fault of specific events and modifies the table counts. \textit{\_\_builtin\_memcmp()} detemines whether the page is missing. When it occurs, the value reference variables such as events and counts will be modified. In our program, we set the specific event to \textit{myprogram}, which can be customized by users.
    \item \textbf{BPF(text = prog)}. The function returns a reference of the code pinned to the kernel. We can get the inkernel data via the reference. In our program, we assign the reference to variable \textit{b} to complete the following work.
    \item \textbf{open\_perf\_buffer(function)}. The function opens a perf ring buffer where stores the data delivered by \textit{BPF\_PERF\_OUTPUT}. At the same time, it calls the function passed in as an argument. In our program, the function is \textit{print\_event()}.
    \item \textbf{print\_event()}. The function receives the data in the perf buffer and prints the information of page fault(number and time).
    \item \textbf{perf\_buffer\_poll()}. The function waits for the data in the perf buffer, preparing for the next data print.
\end{itemize}


We organize these functions. kprobe associate the function \textit{\_\_handle\_mm\_fault()} to the kernel function which records the information of page fault in a table created by \textit{BPF\_HASH()} and \textit{BPF\_PERF\_OUTPUT()}. Then the data will be delivered to the per buffer and printed by the function \textit{print\_event()}.
To execute our program, there are several steps to go through.

\subsubsection{Preparation}

Preparation stage includes compiler and verifier.

\textbf{(1) Compiler.} This compiler is a clang compiler based on LLVM architecture. LLVM provides a platform for eBPF which consists of a subset of C and some helper functions. The subset has several restrictions such as the usage of C function, the types of variables and the size of stack.

\textbf{(2) Verifier.} To ensure the safety of the kernel, the verifier will work after the code is compiled and before it is loaded in the kernel. It checks the source code and ELF of the program in several aspects:
a) The verifier checks the size of the program which should be less than $10^6$ instructions.  
b) The verifier performs Depth-First Search on the eBPF program. If the loops in the program are bounded,  the program can be transformed into Directed Acyclic Graph (DAG).
c) The verifier explores all possible paths from the first instructions by a state machine to check whether the depth is under specified size.
d) The verifier checks the license of functions in the program to make sure it is available and compatible.
e) The verifier performs boundary check on the program to avoid overflow.

\subsubsection{Loading}

An eBPF program should be attached to a hook before it executes. The hook is an interface which can be customized by users. Users can register some programs of several events in the hook. The attached hook of an eBPF program is determined by its type. The eBPF program executes and traces the target data when the registered events happen. In this paper, these events are related to page fault exception.

\subsubsection{Executing}

After preparation and loading stage, the eBPF program is ready to execute and monitor the target data. It works when  the specific event occurs. In this process, maps, helper function and tail calls we mentioned before take effect, in order to complete the data collecting task.

\textbf{(1) Maps}
Shown in figure 3, the eBPF program creates a map which stores the key value of the pid and times of page fault. When executing, the program will modify the data of the map to record the data we need.

The lifetime of map is defined by reference counters(refcnt) in the kernel. When a process creates a new map, the kernel will set map refcnt to 1. The counter calculates the number of the map used by eBPF programs. If refcnt reaches 0, the eBPF program related to the counter will be destroy by the kernel, which means the end of the lifetime of map.

To keep it alive,  maps can be pinned to file system in the kernel. This method is called map pinning, which can be done by several tools such as \textit{bpf()} system call and \textit{libbpf} . 

\textbf{(2) Helper functions}
The eBPF program can call the helper functions in the kernel to perform corresponding tasks, which is one of the differences between eBPF and cBPF. Those tasks are as follow:
a) \textbf{Interacting}. The helper function can interact with the context of hook and with the structure in the kernel such as maps.
b) \textbf{Modifying}. The helper function can modify the data in the process.
c) \textbf{Printing}. The helper function can print the target data to help user monitor the state of program executing.

Declarations of helper functions are mostly in \textit{bpf\_helpers.h}. During compiling process, the compiler only needs to compile the head file consists of the signature of these functions instead of their .c files, for the implementation is provided in the kernel. 

\subsection{Design of Prediction Model}

LightGBM is an efficient framework to implement GBDT algorithnm. GBDT is an outstanding model based on boosting. In boosting algorithnm, in order to improve the effect of learning model and remedy the mistakes of the previous model, the input of \textbf{$y_i$} includes not only the input sequence \textbf{$x$}  but also the training error of last trained model \textbf{$y_{i-1}$}. It can be expressed as the function:
$$H(x) = \sum_{i=1}^N\alpha_{i}h_{i}(x)$$

Where $h_{i}(x)$ is the individual classifier trained by subset of the data. at is the weight of corresponding error. The expression of $\alpha_i$ is:

$$\alpha_{i} = \frac{1}{2}ln\frac{1-\epsilon_i}{\epsilon_i}$$

where $\epsilon_{i}$ is:
$$\epsilon_{i} = P_{x \in D_i}(h_{i}(x) \neq y)$$

The errors of previous model can be remedied via weight addition.
LightGBM optimizes GBDT based on histogram algorithm, which can discretize continuous values into k integers and create a histogram with width k. When processing continuous values, the values are mapped to the histogram. Then search for the best split via the histogram. LightGBM also easily gets a few leaf nodes by subtracting two nodes to make difference with histogram.
In this paper, we get the trained model by LightGBM. The input of model is the number of page fault and the time when a page fault occurs.

\section{Experiments} 
\label{sec:Proposed Solution}
\subsection{Experimental Setup}
The experimental environment of the present paper is based on Linux 5.13.0 and we code eBPF programs with the help of bcc v0.23.0 using Python 3.7. We train our model via LightGBM 3.3.2.

\subsection{Data Collection}
We make use of eBPF programs inserted into kernel mode to hook page faults to collect the times of page faults of our test programs.
    \subsubsection{Data Record} 
    \begin{itemize}
        \item {\textbf{Hash Table Establishment}}   
        To record the times of page faults of observed processes, we build up a hash table named \textit{counts} by the helper function $BPF\_HASH()$, every entry of which the key is the pid of process and the according value is the times of page faults. We build up another hash table called 
        \textit{frq} to record the frequency when page faults occur. When it has reached the threshold $frequency$, which will be dynamically adjusted in real cases, data will be submitted to the perf ring buffer which will be discussed later. If a sample is successfully printed to storage files,  \textit{counts} will be cleared.
        \item {\textbf{Time record}} 
        We define a global variable, \textit{begin\_time}, to monitor the time when the eBPF program executes, which will be later used to calculate the time takes to collect each data sample.
        \item {\textbf{System Call Hook}} 
         We use $kprobe$, a kind of eBPF program, to hook the system call function,  \textit{\_\_handle\_{mm}\_{fault}}, i.e.function $int\  kprobe\_handle\_mm\_fualt(struct pt\_regs *ctx)$ Whenever the system call is called by processes, the program will be trapped into the preset eBPF program for data collection first before handling by the handle functions.
         \item {\textbf{Data Operation inside eBPF Program}} 
        When trapped into the preset eBPF program, the process id and the name of the process will be captured to check if it is the observed process. For a yes, value of the times of page fault in the hash table \textit{counts} will automatically increment. Simultaneously, value of \textit{frq} will increment to record the frequency of page faults. 
    \end{itemize}
    \subsubsection{Data Interaction}
    \begin{itemize}
        \item {\textbf{Perf Ring Buffer Establishment}} 
        To achieve the goal of interacting data from kernel space to user space, we utilize {a BPF table to push out data via a perf ring buffer} called \textit{event} by calling helper function $BPF\_PERF\_OUTPUT()$.
        \item {\textbf{Output Data}} Kernel on Linux offers numerous helper functions to operate perf ring buffers. In this paper, we use {helper functions} $perf\_{buffer}\_poll()$ and $open\_perf\_{buffer}$ to output data.
            Every time the $perf\_{buffer}\_poll()$ is called, the callback function attached to each entry of $open\_perf\_{buffer}$ will be called. 
            We rewrite the callback function renamed \textit{print\_event} to output data. Inside the function \textit{print\_event}, we left out values less than $min\_value$ in consideration of data stability, then output qualified data with timestamp to the user space redirecting to storage files.
    \end{itemize} 
    
    \subsubsection{Approximate WSS Groundtruth}
    In this paper, we utilize the tool offered by \cite{wss} to measure working set size(WSS) in the same time duration when collecting page faults. The target test program designs an integer array with optional length, raging from $[2^{7},2^{22}]$, implementing assignment of each element, in an infinite loop. The program never exits until the running time has reached the time threshold we set to guarantee the consumed memory is endurable to Linux. 
    \\ \indent We have collected more than $20$ data sets, most of which contain samples more than $1200$.%
    \ Marking the value of approximated WSS estimated as the label of times of page faults, prediction model can be trained to estimate future working set size with times of page faults, which will be discussed below.
    
\subsection{Model Establishment}

\subsubsection{Data Preprocessing} 
    To ensure the robustness and stability of our model, we eliminate abnormal data from data collected.
    Moreover, data collection offers times of page faults in the test program during a certain time. We need to calculate the time interval of each period of time to better train our model according to the formula:\ $\Delta t_i = t_{i+1}-t_i$
    Furthermore, we implement data normalization to improve the convergence rate and accuracy of our model. Normalization is usually utilized when there is multi-index evaluation, considering the different nature of evaluation index with different dimensions and orders of magnitude. Dimensions of indexes are different, if directly use the raw index to evaluate, index of high numerical value will play as the primary role in comprehensive evaluation, which weakens the effect of the other indexes with low numerical value. In our paper, two indexes are used.\ They are time interval and times of page faults. Therefore, Normalization used to map range of data collected to the range of $[0, 1]$. The normalization expression is:
    $$X_{standard}= \frac{X-X_{min}}{X_{max}-X_{min}}$$ In this paper, normalization is accomplished by function $MinMaxScaler()$ offered by Python 3.7.
    
\begin{figure*}[htbp]
    \centering
    \subfloat[RMSE of Learning\_Rate]{
        \includegraphics[width=0.25\textwidth]{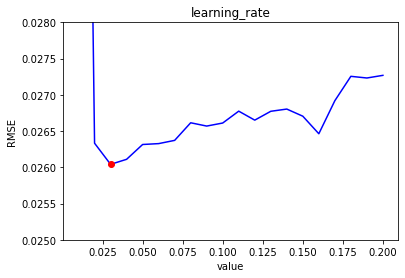}
        \label{fig-my1}
    }
    \subfloat[RMSE of Max\_Depth]{
        \includegraphics[width=0.25\textwidth]{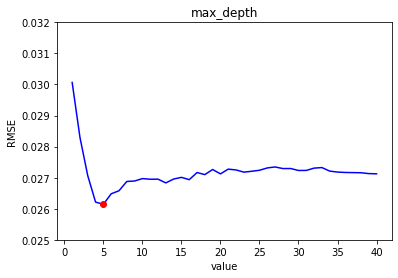}
        \label{fig-my2}
    }
    \subfloat[RMSE of Min\_Child\_Sample]{
        \includegraphics[width=0.25\textwidth]{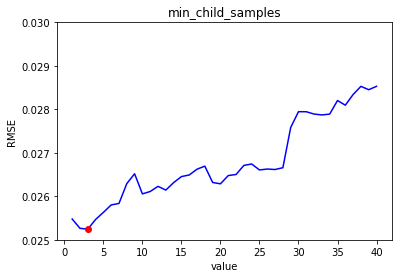}
        \label{fig-my3}
    }
    
    \subfloat[RMSE of Num\_Leaves]{
        \includegraphics[width=0.25\textwidth]{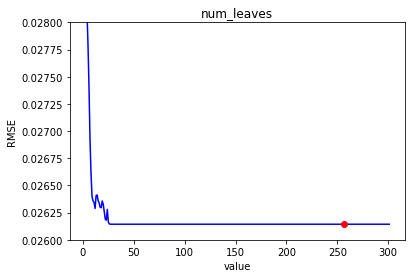}
        \label{fig-my4}
    }
    \subfloat[RMSE of N\_Estimators]{
        \includegraphics[width=0.25\textwidth]{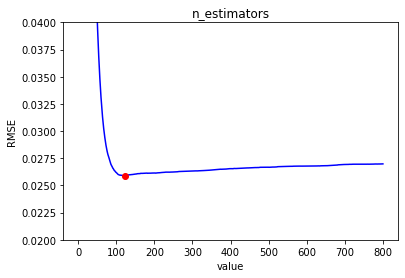}
        \label{fig-my5}
    }
    \subfloat[RMSE of Colsample\_Bytree]{
        \includegraphics[width=0.25\textwidth]{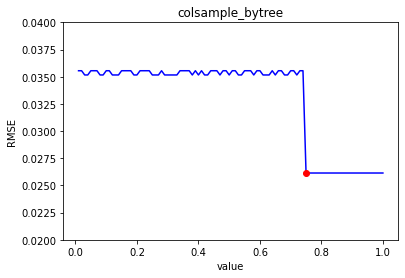}
        \label{fig-my6}
    }
    \caption{Results of hyper-parameter analysis}
    \label{fig:my_label}
\end{figure*}

\subsubsection{Sensitivity Analysis}
LightGBM algorithm offers optional parameters\cite{lightdocs} to train and optimize models, and we have analyzed many of them to find out hyper-parameters. They are listed as follows:
\begin{itemize} 
    \item \textbf{learning\_rate}:\ It is used to control the progress of deep learning models. Its value should be bigger than zero.
    \item \textbf{max\_depth}: The max depth of the decision trees. It is used to limit the depth of a decision tree. Its value should be a positive integer.
    \item \textbf{num\_leaves}:\ The max number of leaves of one decision tree. The value should be a positive integer.
    \item \textbf{n\_estimators}:\ In this paper, we make use of boosting algorithm. The parameter denotes the number of boosting iterations. The value should be a positive integer.
    \item \textbf{min\_child\_samples}:\ Minimal number of data in one leaf. The value should be a positive integer.
    \item \textbf{random\_state}: It is used to generate other seeds. The seed is unused by default due to its low priority comparing to other seeds. It is set to a random value within range $[0,10^6)$.
    \item \textbf{colsample\_bytree}: The percentage of feature selection. The default value is 1, meaning all the features will be selected. Set less than 1, partial features will be chosen at each tree node. Its value should be bigger than zero, less than or equal to one.
    \item \textbf{subsample}: It is used for random selection of data without resampling.The value shold be bigger than zero, less than or equal to one.
\end{itemize}


\indent We make an attempt to find out the optimal interval and suboptimal interval of each hyper-parameter by running the program on NNI\cite{nnidocs}, i.e. Neural Network Intelligence, a toolbox offered by Microsoft for automatic parameters commissioning. RMSE figures of each hyper-parameter are demonstrated from Figure \ref{fig-my1}-\ref{fig-my6}.
From Figure \ref{fig-my1}-\ref{fig-my6}, it's easy to draw the conclusion that nearly each figure presents a smile curve, offering an optimal interval and a suboptimal interval. However, the curve of subsample shows that the value of it makes no impact on RMSE, therefore, its figure is negligible. \\
\indent Optimal interval and suboptimal interval of each hyper-parameter based on each RMSE figure are listed on Table ~\ref{tab2}. 
For learning\_rate, a larger value make it hard for our model to shrink, obtaining a local optimal result rather than a global optimal result. While a smaller value causes a large time overhead to train models. 
 For max\_depth, a larger value brings an over-fitting problem while a smaller value brings an under-fitting problem to our model, making the prediction value inaccurate. For n\_estimators, a larger value takes more time to train our model, while a smaller value causes inaccuracy. For min\_child\_samples, a suitable value makes it feasible to prevent over-fitting. Num\_leaves determines the complexity of a tree in some aspects,a smaller one brings under-fitting while a larger one causes over-fitting. And for colsample\_bytree, it can be used to accelerate the speed of training, a suitable value can also deal with over-fitting problem.
 As we can imagine, inappropriate values of hyper-parameter will bring disastrous blow to the model as well as the whole work. When the value of each hyper-parameter are selected within its optimal interval, the prediction results will be the best with few underestimated and overestimated values, while within its suboptimal interval, some abnormal peak values may occur. The worse situation  As we can imagine, inappropriate values of hyper-parameter will bring disastrous blow to the model as well as the whole work. When the value of each hyper-parameter are selected within its optimal interval, the prediction results will be the best with few underestimated and overestimated values, while within its suboptimal interval, some abnormal peak values may occur. The worse situation occurs when selecting out of optimal intervals or suboptimal intervals, that most of the predicted values seriously deviate from true values, model oscillates dramatically.
\begin{table}[htbp]
\caption{Optimal and Suboptimal Intervals}
\setlength{\tabcolsep}{2mm}
\begin{center}
\begin{tabular}{|c|c|c|} 
\hline 
\cline{1-3} 
\textbf{Parameter} & \textbf{Optimal Interval}& \textbf{Suboptimal Interval}\\
\hline
\textbf{learning\_rate} & $[0.0245,0.0425]$ &$[0.023,0.0245)$\\ & &  $\cup[0.0425,0.073]$\\ 
\hline
\textbf{max\_depth} & $\{4,5\}$ &$\{3,6,7\}$\\
\hline
\textbf{num\_leaves} & $[25,300]$ &$[8,23]$\\
\hline
\textbf{n\_estimators} &$[100,200]$ &$\{99\}\cup [201,310]$\\
\hline
\textbf{min\_child\_samples} & \{2,3\} &\{1,4\}\\
\hline
\textbf{colsample\_bytree} & $[0.75,1]$ &$/$\\
\hline
\end{tabular}
\label{tab2} 
\end{center}
\end{table}
Therefore, the values of hyper-parameters of our trained model are listed below after hundreds of attempts: learning\_rate=$0.04$,max\_depth=$5$,
num\_leaves=$255$,n\_estimators=$200$,min\_child\_samples=$3$,
colsample\_bytree=$0.999$,subsample=$0.57$. These hyper-parameters guarantee the high accuracy of our model.

\begin{figure*}
    \centering
    \subfloat[Output of Test 1]{
        \includegraphics[width=0.22\textwidth]{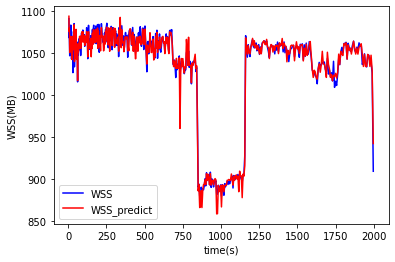}
        \label{fig+2}
    }
    \subfloat[Output of Test 2]{
        \includegraphics[width=0.22\textwidth]{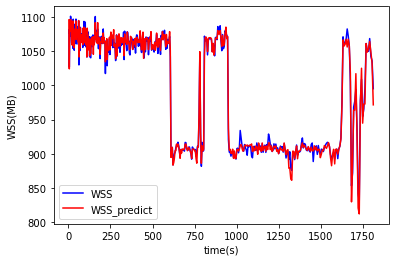}
        \label{fig+3}
    }
    \subfloat[Output of Test 3]{
        \includegraphics[width=0.22\textwidth]{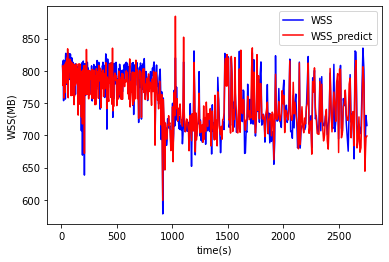}
        \label{fig+4}
    }
    \subfloat[Output of Test 4]{
        \includegraphics[width=0.22\textwidth]{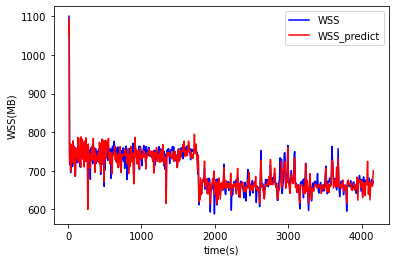}
        \label{fig+5}
    }
    \caption{Performance of WSS estimation among 4 datasets}
    \label{fig:my_label}
\end{figure*}

\subsubsection{Model Training}
To train our model, we classify the data collected into six sets after preprocessing. They are $train\_X$, $train\_Y$, $valid\_X$, $valid\_Y$. $test\_X$ and $test\_Y$. More specifically, they are the training sets, validation sets and test sets. Models learn from the training data and improve and optimize themselves for better prediction results according to the results when inputting validation sets. The size of each set is manually adjusted to optimal our model. The regression model is trained via LightGBM. Hyper-parameters to set the model have been discussed above.

\subsection{WSS Estimation}
\begin{figure}[htbp]
\centerline{\includegraphics[width=0.8\columnwidth]{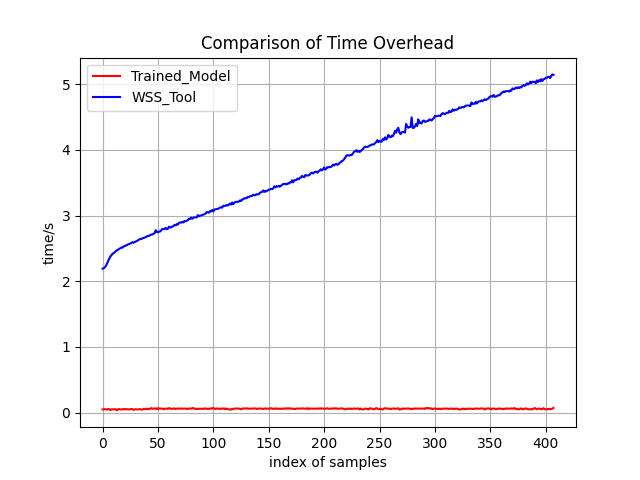}}
\caption{Comparison of Time Overhead}
\label{fig++}
\end{figure}

To verify the accuracy of our work, we make use of 4 test data sets collected from our test programs to test our model.\ Number of samples of each test data set is $444$, $458$, $561$ and $621$. Moreover, we use test programs to simulate the process that a specific process requests memory from the operating system as what has been introduced before. The size of array created by each test program is $200$, $1024$, $1024$ and $25$, ranging from small memory to large chunk of memory. Results can be seen on Figure \ref{fig+2} - \ref{fig+5}. Statistically, RMSE for each test data set is presented on Table \ref{tab1}. With average RMSE equals to $0.0744$, the accuracy of our model is extremely high.\\
\begin{table}[htbp]
\caption{RMSE of Test Data Sets}
\begin{center}
\begin{tabular}{|c|c|c|c|c|c|}
\hline
\textbf{}&
\multicolumn{4}{|c|}{\textbf{Index of Test Data Sets}} \\
\cline{2-5} 
\textbf{} & \textbf{\textit{Data Set 1}}& \textbf{\textit{Data Set 2}}& \textbf{\textit{Data Set 3}}&\textbf{\textit{Data Set 4}} \\
\hline
Size$^{\mathrm{a}}$& 444&458 & 561&621 \\
\hline
RMSE& 0.0014&0.0873 & 0.0291&0.1799 \\
\hline
\multicolumn{3}{l}{$^{\mathrm{a}}$Number of samples in a data set.}
\end{tabular}
\label{tab1}
\end{center}
\end{table}
\indent Furthermore, when it comes to time overhead to estimate working set size, our model shows an overwhelming superiority when compared to WSS tool\cite{wss}. As what is shown in Figure \ref{fig++}, running in our test programs, WSS tool takes a average time, $3.785$s, to output its estimation while our eBPF program takes only $0.0581$s as average time to output results. Nearly $\frac{1}{65}$ of the time WSS tool spent, our model decreases the time overhead greatly.

\section{\textbf{Related Work}}
\label{sec:Related Work}
Multiple papers have committed to estimating working set size with low overhead and high accuracy.
Reuse distance method to construct MRCs is the most commonly introduced method.To construct a MRC of hugepage, \cite{DBLP:conf/ispa/HuBSLWW18} puts forward its proposition to adapt a dynamic hot set to lower the number of page fault of hugepage. Hugepages maintaining in the hot set denotes that their reuse distances are small, thus ruling them out when tracking memory accesses to calculate reuse distance. To keep the high accuracy of MRC, authors propose a theoretical heuristic algorithm to calculate and add the filtered reuse distance back. 
\cite{DBLP:conf/fast/WaldspurgerPGA15} introduces an algorithm called SHARD, constructing online MRCs with uniform randomized spatial sampling. Authors put forward two versions of SHARD, the first one focuses on fixed sampling rate and the second one focuses on a fixed size sample set $S$, a prior queue to store information of referenced pages.It uses hash values of referenced pages to decide whether the page will be sampled or not.The difference occurs when page is inserted to $S$. The second version will evict pages with the largest hash value if $S$ is full.It takes efforts to lower the space overhead and time overhead due to memory accesses.
\\
\indent Taking advantage of internal counters is also a preferable method. 
\cite{DBLP:conf/icac/ChiangLC13}\cite{DBLP:journals/access/HarbyFA19} utilize internal counter $Committed\_AS$ on Linux.
\cite{DBLP:conf/icac/ChiangLC13} designs a Finite State Machine to dynamically detect $Committed\_AS$ and modifies the value of it when necessary. It puts forward three states: the Fast state, the CoolDown state and the Slow state. Initially, the FSM initialized to be on the Fast state,remaining on which the value of $Committed\_AS$ will decrease 5\% per second. Whenever page fault occurs, the next state changes into the CoolDown state where a timer limited to 8 seconds is activated.It stays on CoolDown state as long as page fault occurs within 8 seconds. When timer touches zero, the next state drives into Slow state. When the value of $Committed\_AS$ changes, its next state will be Fast state and the working set size estimation will be modified to the exact value of $Committed\_AS$. 
\cite{DBLP:journals/access/HarbyFA19} puts forward another FSM which is similar to the former one, while adding a Recovery state. 
The initialized state is the Fast state, however, whenever $Committed\_AS$ changes, it changes into the Recovery state where the last state is not Recovery state. And when in the Recovery state, once $Committed\_AS$ increases by more than 5\%, it changes into the Fast state, otherwise changes into the Slow state, once $Committed\_AS$ decreases, it changes incrementally until the system becomes steady again. When page fault occurs, it changes into the CoolDown state as the same as the former FSM. However, the utilization of FSMs needs to consider more detailed to cover all complex situations.\\
\indent To the best of our knowledge, we are the first work to monitor memory activities with the help of eBPF tools. Our work employs eBPF tools to collect times of page faults and makes use of LightGBM to predict future working set size.
 
\section{\textbf{Conclusion}}  
\label{sec:Conclusion}
In this paper we proposed a novel non-intrusive framework to estimate working set size in a highly efficient way.
To the best of our knowledge, this is the first work that estimate WSS of process with a non-intrusion technique while guaranteeing low overhead.
Furthermore, we preprocess the collected data and train the model by LightGBM to estimate WSS accurately.
Experimental results indicate that our model reduces the overhead by 65.2x compared to previous work.  
Meanwhile,  the average RMSE is only 0.0774,  revealing high estimated accuracy of our model.


\section{Acknowledgment}
The work was supported by Key-Area Research and Development Program of Guangdong Province (No. 2020B010165002).

\bibliographystyle{plain}

\end{document}